\documentclass[aps,prl,twocolumn]{revtex4}

\usepackage{amsmath}
\usepackage{amssymb}
\usepackage{natbib}
\usepackage{graphicx}
\usepackage{bm}

\usepackage{subfigure}
\usepackage{float}
\usepackage{epsfig}
\usepackage{hyperref}
\hypersetup{
colorlinks=true,final=true,
        linkcolor=blue,
        citecolor=blue,
        filecolor=blue,
        urlcolor=blue,
}

\def\be{\begin{equation}}
\def\ee{\end{equation}}
\def\ba{\begin{eqnarray}}
\def\ea{\end{eqnarray}}

\def\ve{\varepsilon}

\def\EE{\textbf{E}}

\begin{document}

\title{Chiral magnetic effect of hot electrons}
\author{S. Nandy}
\author{D. A. Pesin}
\affiliation{Department of Physics, University of Virginia, Charlottesville, VA 22904 USA}

\begin{abstract}
We propose a way to observe the chiral magnetic effect in non-centrosymmetric Weyl semimetals
under the action of strong electric field, via the non-linear part of their I-V characteristic
that is odd in the external magnetic field, or odd-in-magnetic field voltages in electrically open circuits. This effect relies on valley-selective heating in such
materials, which in general leads to nonequilibrium valley population imbalances. In the presence of an external magnetic field, such a valley-imbalanced Weyl semimetal will in general develop an electric current along the direction of the magnetic field -- the chiral magnetic effect. We also discuss a specific experimental setup to observe the chiral magnetic effect of hot electrons.
\end{abstract}
\maketitle

{\color{blue}{\em Introduction}} -- In the area of three-dimensional topological systems, the theoretical predictions and experimental discoveries of Weyl semimetals (WSMs) have led to an explosion of activities due to the intriguing topological properties of these materials. Weyl semimetals appear as topologically nontrivial conductors where the spin-nondegenerate valence and conduction bands touch at isolated points, the so-called ``Weyl nodes", which act as the sources and sinks of the Berry curvature, which is an analog of the magnetic field in the momentum space~\cite{Murakami07,Murak07, WanVishwanath2011,Ran2011,BurkovTopoNodal,
WeylMultiLayer,Xu2011}. In these systems, which violate spatial inversion symmetry and/or time reversal (TR) symmetry, Weyl points of opposite chirality come in pairs due to a no-go theorem by Nielsen and Ninomiya~\cite{Nielsen_1981, Nielsen1983e}.

In this work, we focus on one of WSM signature transport properties: the chiral magnetic effect (CME). The CME describes the generation of an electric current parallel to an applied magnetic field ($\bm B$) induced by the chirality imbalance~\cite{Nielsen1983e, Fukushima_2008, Son_2013, Son2012, Vilenkin, Alekseev1998, Kharzeev2009}. In the context of WSMs, the corresponding current can be written as~\cite{Franz2013, Chen2013a, Zhou2013, Son2012, Vilenkin, Alekseev1998, Kharzeev2009}
\begin{eqnarray}\label{CM_G}
\bm{J}_{cme}=\frac{e^2}{4\pi^2\hbar^2}\sum_{w}\chi_w \mu_w \bm{B}
\end{eqnarray}
where $w$ is the valley index, $\chi_w$ and $\mu_w$ are the monopole charge and chemical potential of the $w^{th}$ valley respectively. Note that all chemical potentials are counted from a common origin. In what follows we will use $\zeta_w\equiv \mu_w-E_w$ to denote the doping level of a node counted from the energy of the band touching.

The possibility to observe the CME current, Eq.~(\ref{CM_G}), relies on one's ability to drive valleys of opposite chirality out of equilibrium with each other. Indeed, it is clear that for $\mu_w=\mu$, the Berry-neutrality condition $\sum_w \chi_w=0$ ensures that the CME current vanishes. In WSMs, the imbalance between valleys of opposite chirality can be achieved via the chiral anomaly. This route was taken in proposals to measure the CME in crystals via classical negative magnetoresistance~\cite{Nielsen1983e,SonSpivak2013}, or nonlocal voltages~\cite{Parameswaran2014}. The key feature of the anomaly-based proposals to uncover the CME is the fact that the external magnetic field is used to both generate valley chemical potential imbalances, and to convert them into a CME current. Therefore, the resultant signals are \textit{even} in the magnetic field. While it is still possible to measure them~\cite{zhang2017room,ong2018anomaly,deboer2019nonlocal}, great care must be taken to distinguish the topology-related effects from mundane Ohmic physics~\cite{Hassinger2016anomaly,ong2018anomaly}

In this Letter, we present a new way to observe the chiral magnetic effect in non-centrosymmetric Weyl semimetals under the action of strong electric fields, via the non-linear part of the I-V  characteristic that is odd in the external magnetic field. In this approach, the chiral imbalance is generated by valley-dependent heating which occurs either due to anisotropy of a crystal\cite{hart1970fluctuations}, or its gyrotropy. We show that valley-selective Joule heating leads to hot carrier redistribution among Weyl nodes with opposite chiralities. When subject to an external magnetic field, such a valley-imbalanced Weyl semimetal will in general develop an electric current along the direction of the magnetic field. We call the appearance of such a current the CME of hot electrons. 

{\color{blue}{\em Hot electrons in WSMs}} -- We view a Weyl semimetal as a collection of anisotropic Weyl nodes, which are labeled with index $w$ and are described by the Weyl Hamiltonian:
\begin{equation}
{ H_{w}(\bm{k}})=\chi_w{\hbar}v^w_{ab}k_a \Sigma_b+E_w
\label{H_linear}
\end{equation}
where $v_{ab}$ with Cartesian indices $a,b$ is the velocity tensor with positive determinant; $\chi_w$ is the chirality associated with the Weyl node; $\Sigma_a$ is the $a$'th Pauli matrix; $E_w$ describes the position of the Weyl node in energy space.

With each anisotropic Weyl point described by Hamiltonian~\eqref{H_linear} we can associate a conductivity tensor $\sigma^w_{ab}$, which is responsible for the valley-specific Joule heating
\begin{align}\label{eq:Joule}
  P^{\rm Joule}_w=\sigma^w_{ab}E_aE_b,
\end{align}
$\EE$ being the electric field applied to the crystal. Balancing the rate of heat production, Eq.~\eqref{eq:Joule}, against the rate of energy transfer into the phonon subsystem determines the steady-state temperature of a node.

In what follows we describe the relaxation processes of hot carriers in Weyl semimetals~\cite{fiete2015cooling}. The relevant scattering mechanisms and the corresponding typical time scales are: intravalley impurity scattering, $\tau$; intravalley electron-electron scattering, $\tau_{ee}$; intravalley electron-phonon scattering, $\tau_{ph}$; intervalley scattering, $\tau_v$. Here we assume the following hierarchy of the relaxation times: $\tau\ll\tau_{ee}\ll\tau_{ph}\ll\tau_v$. The $\tau_{ee}\ll\tau_{ph}$ inequality holds for temperatures that are not too low, see below. We also assume the Fermi-liquid regime to hold.

The above hierarchy of times allows us to simplify the problem by avoiding explicit consideration of the two fastest processes. Of these, the impurity intravalley scattering determines the odd-in-momentum part of the electron distribution function, and the conductivity of a valley. The intravalley electron-electron scattering brings the energy-dependent part of the distribution function to a quasi-quasiequilibrium form with valley-specific values of electronic temperature and chemical potential.

The two slower processes that are key for our purposes are the intravalley electron-phonon scattering, and intervalley scattering of charge carriers. Electron-phonon scattering transfers energy out of the electronic subsystem and determines the steady-state value of a node's electronic temperature, $T_w$. The intervalley scattering, regardless of its origin, redistributes carries among Weyl nodes, determining their non-equilibrium chemical potentials, $\mu_w$. We discuss these two processes in what follows.

Starting with the valley temperatures, we note that their steady state values are found from balancing the Joule heating, Eq.~\eqref{eq:Joule}, with the electron energy loss to phonons within each valley (since intervalley energy transfer is a slow process). For a single isotropic valley, the energy loss due to electron-phonon scattering was considered in Ref.~\onlinecite{fiete2015cooling}. The result is most economically expressed using a parameter $\lambda=\frac{k_B^5D^2}{16\pi\hbar^4 v_F \rho v_s^4}$, which involves the deformation potential $D$, crystal mass density $\rho$, speed of sound $v_s$, and the typical Fermi velocity $v_F$. We also introduce the characteristic Bloch-Gruneisen (BG) temperature given by $k_{B}T_{BG}=2\hbar v_{s} k_{F}$, the density of states at the Fermi level, $N(\zeta)=\frac{\zeta^2}{2\pi^2 \hbar^3 v_F^3}$, and suppress the index $w$ in all valley-dependent quantities except the temperature.  The energy loss for a single valley per unit time and unit volume of the crystal is then given by
\begin{eqnarray}
P^{\rm {e-ph}}&&=-N(\zeta)\lambda T_{BG}^4 (T_w-T).
\label{P_tot}
\end{eqnarray}

In a steady state one has $P^{\rm {Joule}}+P^{\rm{e-ph}}=0$, which yields an electronic temperature
\begin{align}\label{eq:valleyT}
  T_w=T+\frac{\sigma_{ab}E_aE_b}{N(\zeta)\lambda T_{BG}^4}.
\end{align}

Before moving to a discussion of intervalley scattering and chemical potential imbalances, we briefly comment on the region of Eq.~\eqref{eq:valleyT} validity. From here on, we switch to the system of units with $\hbar=k_B=1$, since in the expressions below these constants appear in a trivially predictable way. Eq.~\eqref{eq:valleyT} relies on the existence of electronic temperature, and on the temperature being high compared to $T_{BG}$, such that the electron-phonon collisions are quasi-elastic. The first condition requires electron-electron collisions be faster than the electron-phonon ones. The electron-phonon scattering rate is\cite{sarma2015scattering} $\tau_{ph}^{-1}\sim\lambda T T_{BG}^{2}$, while the electron-electron one is $\tau_{ee}^{-1}\sim  T^2/ N_v^2\zeta$, where $N_v$ is the number of valleys in a WSM. We observe that the electron-electron collisions dominate for $T\gtrsim N_v^2\lambda\zeta T_{\rm BG}^2$. For typical numbers, electron-electron collisions dominate for temperatures above a few Kelvin. Since the Bloch-Gr\"uneisen temperature is roughly a Kelvin in typical WSMs\cite{fiete2015cooling}, we see that the temperature regime in which Eq.~\eqref{eq:valleyT} holds is determined by the $\tau_{e-ph}>\tau_{ee}$ condition, while $T>T_{BG}$ is a weaker one.

Turning to the intervalley scattering, we assume that it happens mainly due to impurity scattering. This is a good approximation at low temperatures, but also at temperatures large compared to the Bloch-Gr\"uneisen temperature corresponding to the typical intervalley momentum transfer, in which case the electron-phonon scattering is quasi-elastic. Hence we expect it to qualitatively describe the physical situation at all relevant temperatures.

We describe the intervalley impurity scattering with a scattering rate $\Gamma_{ww'}(\ve)$, which sets the rate of transitions from valley $w'$ to valley $w$ per unit energy range, per unit volume. We neglect ``skew" intervalley scattering, setting $\Gamma_{ww'}=\Gamma_{w'w}$. Under these assumptions, the rate of change of the particle density in valley $w$, $n_w$, due to the intervalley scattering is given by
\begin{align}\label{eq:ndot}
  \dot n_w=-\sum_{w'}\int d\ve \Gamma_{ww'}(\ve)(f_w(\ve)-f_{w'}(\ve))
\end{align}
Here $f_w(\ve)$ is the angle-averaged distribution function of carriers in valley $w$, which is only a function of the carrier's energy. The steady-state chemical potentials are from $\dot n_w=0$. Recalling that for $\tau_{ee}\ll \tau_{ph}$ the distribution function $f_w(\ve)$ has a quasiequilibrium form with a valley-dependent chemical potential $\mu_w$ and temperature $T_w$, and applying Sommerfeld expansion to Eq.~\eqref{eq:ndot}, we obtain a system of equations for the valley chemical potentials:
\begin{align}\label{eq:valleymu}
  \sum_{w'}\Gamma_{ww'}(\mu_w-\mu_{w'})+\frac{\pi^2\Gamma'_{ww'}}{6}(T_w^2-T_{w'}^2)=0.
\end{align}
At most $N_v-1$ of these equations are linearly independent because of particle conservation by intervalley scattering. They are sufficient to determine valley chemical potential differences driven by valley-dependent temperatures of Eq.~\eqref{eq:valleyT}. Therefore, Eq.~\eqref{eq:valleymu} fully describes the CME in the system of hot electrons. Its validity relies on the intervalley scattering being the slowest relaxation process.

{\color{blue}{\em CME of hot electrons in simple models}} -- Below we consider the CME current in two  simple models of a WSM, in which the considerations are effectively reduced to just two inequivalent valleys.

\begin{figure}[t]
\begin{center}
\includegraphics[width=3in]{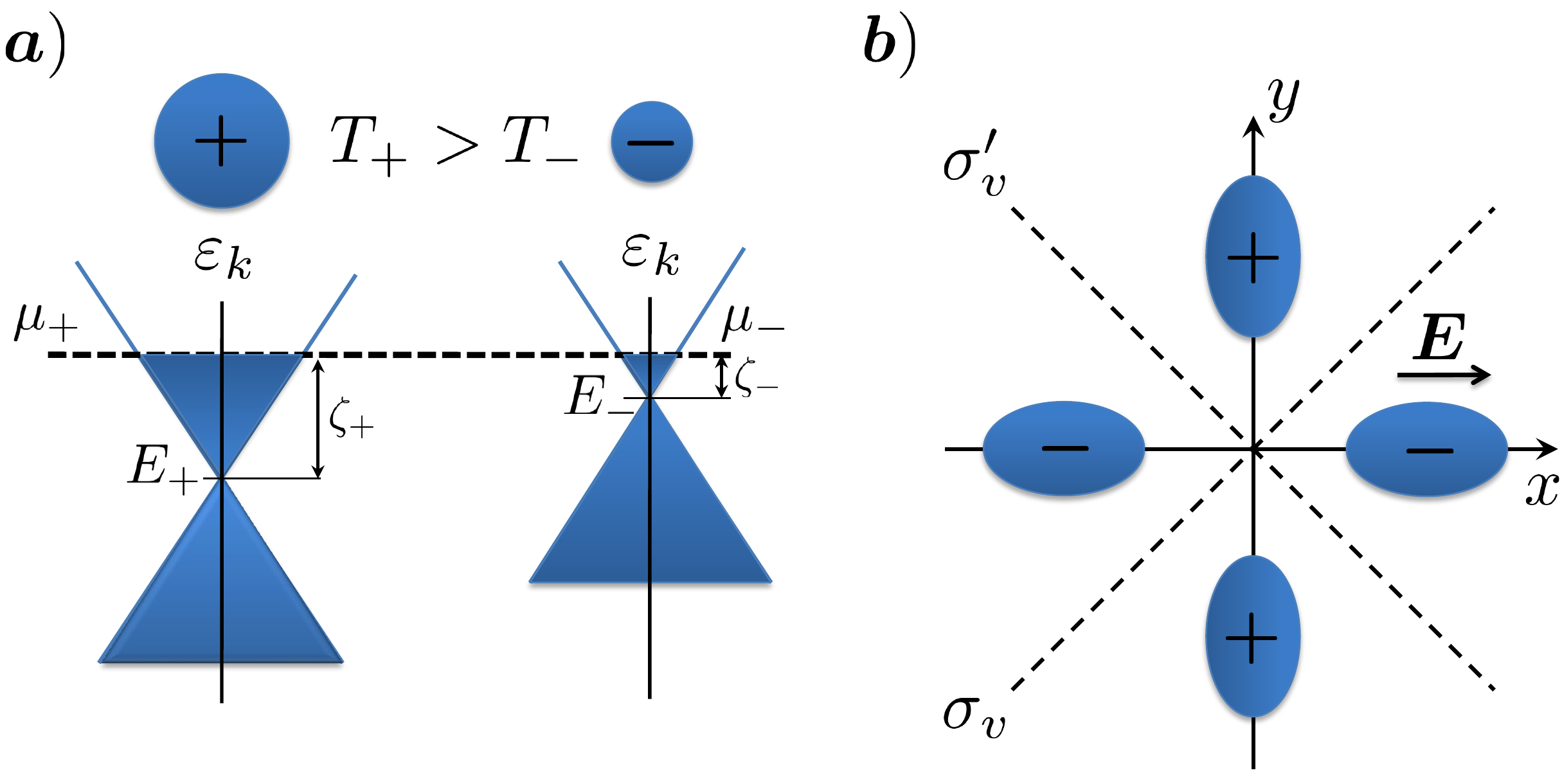}
\caption{(Color online) (a) Local in momentum space band structure of a Weyl semimetal with two Weyl nodes located at different energies, $E_{+}< E_-$, and different positions in momentum space. $``+"$ and $``-"$ represent the chiralities of the Weyl nodes. The size of the Fermi surfaces are different for $``+"$ and $``-"$ valleys, as measured by $\zeta_+>\zeta_-$, where $\zeta_{\pm}$ represent the chemical potentials for $\pm$ valleys counted from the nodal energies. In the presence of an electric field, the valley with the larger conductivity has higher temperature. (b) A simple model of a TR-invariant WSM with the $C_{2v}$ point group and 4 nodes related by TR and mirror symmetries. The $C_2$ axis is perpendicular to the plane of the figure. The Fermi surfaces of the mirror-related nodes are assumed to be anisotropic, and are shown with ellipses.  In the presence of a strong electric field $\bm E$ applied along the fast direction of one of the valleys, this valley's temperature is higher than that of its mirror-reflection partner, which has the electric field point along its slow direction.}
\label{CME_1}
\end{center}
\end{figure}

First, we consider a WSM with just two Weyl nodes, which are located at different energies, Fig.~\ref{CME_1}a. This is a minimal model of a gyrotropic (no mirror symmetries) WSM with broken time-reversal symmetry\cite{noteTR}.

We assume that the valleys are isotropic, such that the conductivity tensor in Eq.~\eqref{eq:valleyT} must be replaced according to $\sigma_{ab}\to\sigma_w\delta_{ab}$, $\delta_{ab}$ being the Kronecker symbol. We will use $w=\pm$ to label the valleys according the their chiralities. For definiteness, let us assume that the valley with positive chirality has a larger Fermi surface due to the corresponding nodal point being lower in energy, $E_+<E_-$ in Eq.~\eqref{H_linear}, while the rest of their microscopic parameters are the same. This implies that $\sigma_+>\sigma_-$. Since there are only two valleys, we can drop the subscript on the transition rates, $\Gamma_{ww'}\to \Gamma$.

From Eqs.~\eqref{eq:valleyT} and \eqref{eq:valleymu} we obtain the difference between the chemical potentials of the two valleys, and use it to calculate the CME current, Eq.~\eqref{CM_G}. As a result, we obtain a non-linear contribution to the I-V characteristics of the WSM, which is odd in the external magnetic field, $\bm j_{cme}=\alpha_{g}E^2\bm B$, where
\begin{align}\label{eq:gammag}
  \alpha_{g}=-\frac{e^2}{12}\frac{\Gamma'}{\Gamma}T\left[\frac{\sigma}{N(\mu)\lambda T_{BG}^4}\right]_{\pm}.
\end{align}
Here we assumed moderate electric fields, such that $T_+^2-T_-^2\approx 2T (T_+-T_-)$, and the symbol $[\ldots]_{\pm}$ denotes the difference of the entire expression inside the bracket evaluated in the $``+"$ and $``-"$ valleys. In this model, the breaking of inversion symmetry required for the appearance of the CME current of hot electron is signalled by $[\ldots]_\pm\neq 0$.

Let us now consider a minimal model of a TR-invariant WSM with the $C_{2v}$ point group, which includes four nodes, see Fig.~\ref{CME_1}b. In this model, the valleys related by the TR symmetry are identical, hence have the same chirality, transport characteristics, temperatures and chemical potentials. Valleys that are related by the mirror symmetry have opposite chiralities; their conductivity tensors are essentially one and the same tensor, but with respect to different (rotated by $\pi/2$ around the polar axis) set of axes. For simplicity, we assume that the conductivity tensor is diagonal.

Being symmetry related, the Weyl nodes of the present model are all at the same energy, hence the preceding considerations do not apply directly. That this model nevertheless does exhibit the CME of hot electrons can be easily seen from the following argument. Consider an electric field oriented along the $x$-axis, as shown in Fig.~\ref{CME_1}b. This is a slow direction for the valleys with negative chirality, and is the fast direction for the ones with positivity chirality, see Fig.~\ref{CME_1}b. Therefore, we expect that for this electric field orientation the $``+"$ valleys will have a higher temperature than $``-"$ valleys. According to the preceeding considerations, that will result in electron transfer from the hot to cold valleys, and hence non-zero CME current. It is clear that the sign of the effect will be reversed for the electric field oriented along the $y$-axis, assuming the same orientation of the $\bm B$-field. The effect vanishes for electric fields in the mirror planes of the crystal, since such fields do not break the symmetry between the valleys with opposite chiralities. These considerations show that the CME current in this model is $\bm j_{cme}=\alpha_{c_{2v}} (E_x^2-E_y^2)\bm B$. This is consistent with the symmetry requirements of the $C_{2v}$ group.

We now turn to the quantitative theory of $\alpha_{c_{2v}}$. First, we note that since the valleys with opposite chiralities are related by mirror symmetry, their diagonal conductivity tensors are given by $\sigma_+={\rm{diag}}(\sigma_{xx},\sigma_{yy},\sigma_{zz})$ and $\sigma_-={\rm{diag}}(\sigma_{yy},\sigma_{xx},\sigma_{zz})$, respectively. Their densities of states at the Fermi level are the same, and we also assume that they can be assigned effective Bloch-Gr\"uneisen temperatures, which are also the same by symmetry. Using the above conductivity tensors in the equation~\eqref{eq:valleyT} for the valley temperatures, we obtain
\begin{align}
  T_+-T_-=\frac{\sigma_{xx}-\sigma_{yy}}{N(\mu)\lambda T_{BG}^4}(E_x^2-E_y^2).
\end{align}
As is clear from this equation, and as was explained above, the temperature difference between valleys is driven by valley anisotropy in this case.

In the present model, the intervalley scattering only operates between the mirror-symmetry related valleys of opposite chiralities, since the chemical potentials and temperatures of the TR-related valleys are the same. Hence this four-valley model effectively reduces to a two-valley one, and the considerations  of the previous model of Fig.~\ref{CME_1}a apply. The expression for $\alpha_{c_{2v}}$ ends up being
\begin{equation}\label{eq:alphaC2v}
  \alpha_{c_{2v}}=-\frac{e^2}{12}\frac{\Gamma'}{\Gamma}T\frac{\sigma_{xx}-\sigma_{yy}}{N(\mu)\lambda T_{BG}^4}.
\end{equation}

Equation~\eqref{eq:alphaC2v} allows to estimate the order of magnitude of the CME of hot electrons. We assume that the scattering rate $\Gamma(\ve)$ has a smooth energy dependence on the scale of a typical Fermi energy, $\Gamma'/\Gamma\sim 1/\zeta$, and use typical numbers for a WSM: $\zeta=15\,\rm{meV}$, $v_F=4\times 10^{5}\,\rm{m/s}$, $v_s=2.8\times 10^3\,\rm{m/s}$, $D=20\,\rm{eV}$, $\rho=7\times 10^3\,\rm{kg/m^3}$, mobility $\mu_{tr}=10^5\,\rm{cm^2/Vs}$, and anisotropy of $20\%$. Then at $T=10\,\rm{K}$ we obtain $\tau_{ph}\sim10^{-10}\,\rm{s}$ which is comparable to the typical disorder-induced intervalley scattering times, hence our results apply for $T\gtrsim 10\rm K$ for this hypothetical material (such that $\tau_{ph}<\tau_{v}$). At $T=10\,\rm K$ we get $|\alpha_{c_{2v}}|\approx 10^2\, \rm{T^{-1}V^{-1}\Omega^{-1}}$. This is a very large value of $\alpha_{cme}$, which can grow further with temperature, in an approximately linear fashion. We further discuss this point in the concluding part of the paper.

{\color{blue}{\em General symmetry requirements and candidate materials}} -- The general expression for the CME current of hot electrons, $j_{cme, a}=g_{bc}E_bE_c B_a$, is determined by a symmetric second-rank pseudotensor $g$. Therefore it can exist only in (gyrotropic) crystals with point groups allowing such a tensor. These are the same crystals that show natural optical activity, the symmetry requirements for which are discussed at length in textbooks\cite{Malgrange2014}.

{\color{blue}{\em Discussion}} -- We would like to conclude with discussing the relation of our results to the previous work, and describe an experimental setup to measure the CME of hot electrons. Non-linear transport effects that are odd in magnetic field have a long history in conventional non-centrosymmetric semiconductors~\cite{Ivchenko1983}, macroscopic conducting helices\cite{wyder2001bi}, and chiral carbon nanotubes~\cite{ivchenko2002nanotube}. In the context of WSMs, the most relevant for the present work is Ref.~\cite{Morimoto2016}, which studied the appearance of the magneto-chiral anisotropy in WSM due to the chiral anomaly. In the language of the present paper, that amounts to a non-linear in electric field current that is driven by the chiral anomaly, and has the following form in a WSM with isotropic valleys: $\bm j_{E^2B}=\alpha_{an}(\bm E\cdot\bm B)\bm E$. Under the same conditions, the current studied in the present work is given by $\bm j_{cme}=\alpha_{cme}E^2\bm B$. The most notable difference between these two currents is their dependence on the orientation of the electric and magnetic fields. While the current studied in Ref.~\cite{Morimoto2016} requires that the magnetic field be aligned with the electric field, while the current itself flows along the electric field, the current studied here exists for any mutual orientation of the $\bm E$ and $\bm B$ fields, and flows along the magnetic field. This difference can be used to distinguish between the two effects experimentally, see below.

One can also compare the magnitudes of the two currents using the expressions obtained from the model of Fig.~\ref{CME_1}a, which was also employed in Ref.~\cite{Morimoto2016}. After bringing the results of Ref.~\cite{Morimoto2016} to the present notation, and some simple algebra, we get $\alpha_{cme}/\alpha_{an}\sim \frac{T^2}{T_{BG}^2}\frac{\tau_{ph}}{\tau_v}$, where $\tau_v$ is the intervalley scattering time. Thus the ratio of magnitudes of the two effects contains two factors, the first of which, $T^2/T_{BG}^2$ can be made large, and the other one, $\tau_{ph}/\tau_v$, is typically small. Our estimates show that the two effects are roughly of equal magnitude a temperature of about $10\,\rm K$, above which the CME-driven effect considered here overpowers the chiral anomaly-related one. Both effects are several orders of magnitude stronger than their analogs in conventional materials~\cite{Morimoto2016}. 

The considerations of the preceding paragraph also make it clear that the two--anomaly- and CME-related-- effects have different temperature dependencies. The anomaly-related effect is finite at zero temperature, the corrections at finite temperature going like $T^2/\zeta^2$. Instead, the CME-related effect of the present work is small at small temperatures, but grows with temperature approximately linearly at $T\gtrsim T_{BG}$. 

\begin{figure}[t]
\begin{center}
\includegraphics[width=3in]{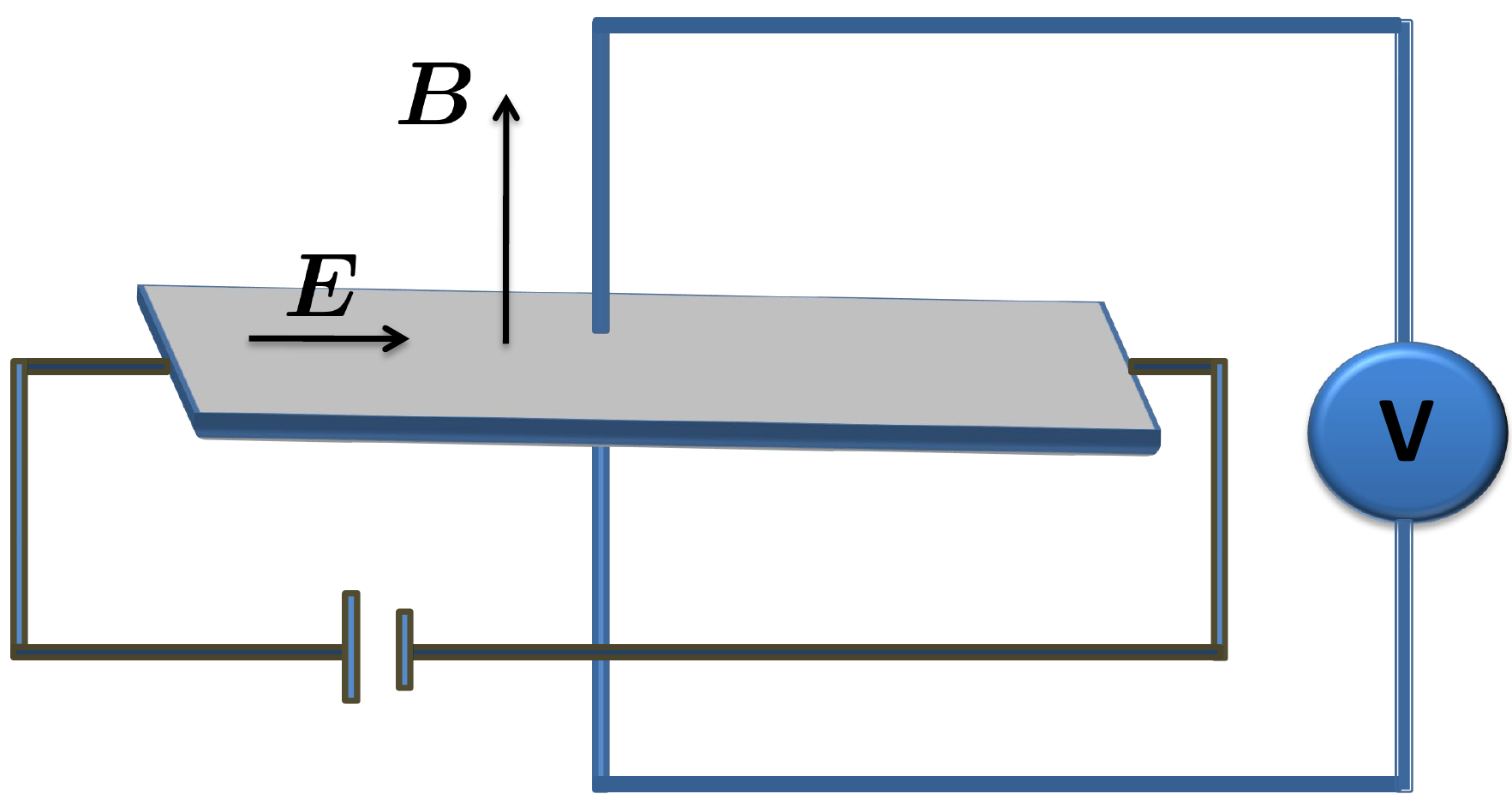}
\caption{(Color online) Illustration of a thin-film CME measurement geometry in the presence of a strong in-plane electric field ($\bm E$). A magnetic field $\bm B$ is applied perpendicular to the plane of the sample. A voltage measured perpendicular to the plane is indicative of the CME of hot electrons.}
\label{CME_setup}
\end{center}
\end{figure}

Finally, we describe a setup to measure the CME of hot electrons, see Fig.~\ref{CME_setup}. Most drastically this effect can manifest itself via odd-in-B open-circuit voltages that vanish without a magnetic field. In the thin-film geometry of Fig.~\ref{CME_setup}, there is ideally no voltage in the direction perpendicular to the in-plane current flow. Upon application of an out-of-plane magnetic field, a voltage drop will develop across the film, whose magnitude is set by the condition that there be no net current in the electrically-open circuit. The corresponding electric field across the film is given by $E_\perp\sim \alpha_{cme} E^2B/\sigma$, where $E$ is the in-plane transport electric field, and $\sigma$ is the relevant conductivity. For $E=10\, \rm{V/m}$ and $B=0.1\,\rm T$, we obtain $E_\perp\sim 0.5\,\rm{V/m}$ for the numbers quoted above for the toy model with $C_{2v}$ symmetry. Since the sign of the effect in general depends on the transport electric field orientation with respect to the crystallographic axes, it appears that the strongest limitation on the observability of the CME signal is put by the requirement that the sample be a single crystal.

\textit{Acknowledgments:} This work was supported by the National Science Foundation Grant No. DMR-1853048.

\bibliography{WSM_references}

\begin{thebibliography}{33}
\bibitem[{not()}]{noteTR}
\bibinfo{note}{In this model with broken TR there must exist a background
  magnetization, $\bm M$, proportional to the momentum-space splitting between
  the Weyl point. Symmetry-wise, there is an allowed term in the current of the
  form $j\propto E^2(\bm B\cdot\bm M)$, which requires broken inversion
  symmetry. Such a contribution is of no interest to us.}
\end{thebibliography}
\bibliographystyle{apsrev}

\end{document}